\documentstyle[aps,amsmath,latexsym,amssymb,epsfig,multicol,graphicx]{revtex}

\begin{document}

\draft

\title{The crossover between lasing and polariton condensation in optical microcavities}

\author{M. H. Szymanska${}^{1}$, P. B. Littlewood${}^{1,2}$}
\address{${}^{1}$Theory of Condensed Matter, Cavendish Laboratory, Cambridge
CB3 0HE, UK\\
${}^{2}$Bell Laboratories, Lucent Technologies, Murray Hill, NJ 07974 USA}

\date{\today}

\maketitle

\begin{abstract}

We study a model of a photon mode dipole-coupled to a medium of
two-level oscillators in a microcavity in the presence of dephasing
processes introduced by coupling to external baths. Decoherence
processes can be classified as pair-breaking or non-pair-breaking in
analogy with magnetic or non-magnetic impurities in
superconductors. In the absence of dephasing, the ground state of the
model is a polariton condensate with a gap in the excitation
spectrum. Increase of the pair-breaking parameter $\gamma$ reduces the
gap, which becomes zero at a critical value $\gamma_{C1}$; for large
$\gamma$, the conventional laser regime is obtained in a
way that demonstrates its close analogy to a gapless superconductor.
In contrast, weak non-pair-breaking processes have no qualitative
effect on the condensate or the existence of a gap, although they lead
to inhomogeneous broadening of the excitations.

\end{abstract}

\begin{multicols}{2}
\narrowtext

There are two apparently distinct phenomena where quantum coherence
has been observed on the macroscopic scale: the laser and Bose
Einstein condensation (BEC) of massive particles.

The laser is a weak-coupling phenomenon: a coherent state of photons
created by stimulated emission from an inverted electronic population
due to the strong pumping. The polarisation of the medium is heavily
damped and the atomic coherence is very much reduced.  A coherent
photon field, oscillating at bare cavity mode frequency, is the only
order parameter in the system \cite{laser}.

Polaritons are strongly coupled modes of light and electronic
excitations. A theoretically constructed
polariton condensate is a mixture of coherent state
of light and coherent state of massive particles in the media. It is
characterised by two order parameters: the coherent polarisation and
the coherent photon field and exhibit a gap in the excitation spectrum
\cite{Paul}.

The simplest models of both the laser and the polariton condensate
start from similar equations of motion, but make different physical
approximations. The laser is assumed to be far from equilibrium, and
decoherence (dephasing) --- arising from collisions, interactions with
phonons, defects or impurities, and manifested by homogeneous
broadening ($1/T_2$) of the electronic excitations --- is imposed by
hand. In contrast, the polariton condensate has been assumed to be in
thermal equilibrium, and dephasing phenomena neglected.

\begin{figure}
     \begin{center} \leavevmode \epsfxsize=8.8cm
     \epsfbox{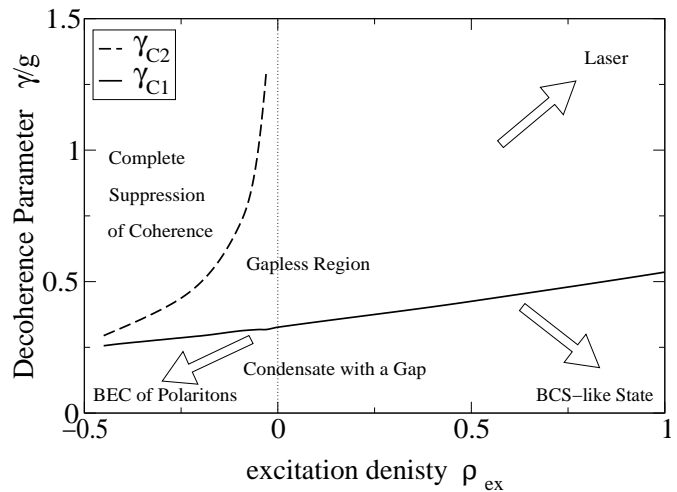}
     \end{center} \caption{Phase diagram. $\rho_{ex}$ is a density of
     total excitations in the microcavity, which is a sum of
     photons and electronic excitations. In this terminology the
     minimum $\rho_{ex}=-0.5$ corresponds to no photons and no
     electronic excitations in the system, while 0.5 would
     correspond to the maximum of electronic excitations in the
     absence of photons. The phase boundaries $\gamma_{C1}$ and
     $\gamma_{C2}$ are marked in Fig. \ref{fig:phot}.  }
\label{fig:phase}
\end{figure}

In this letter, we show that a consistent inclusion of decoherence
phenomena drives a crossover from the polariton condensate to the
lasing regime. Using a model of two-level oscillators coupled to a
degenerate optical cavity mode with coupling strength $g$, and with
further coupling of the internal degrees of freedom to external baths
(defined in the Hamiltonian below), we find the generic phase diagram
of Fig. \ref{fig:phase}, where a single pair-breaking parameter
$\gamma$ encapsulates the strength of electronic dephasing. When the
decoherence is weak, we find a strong-coupling regime with a polariton
condensate and a gap in the excitation spectrum.  The gap is robust
against (weak) dephasing processes, a result that can be obtained only
by the correct self-consistent treatment of decoherence processes. For
larger decoherence, the gap is suppressed to zero at a critical value
$\gamma_{C1}$ in a similar way to the effect of pair-breaking
processes in a BCS superconductor, and the coherent polarisation is
reduced. For low excitation densities coherence is completely
suppressed at a higher value $\gamma_{C2}$.  In the limit of high
excitation densities (photon dominated) and large decoherence, we
recover the traditional model of the laser.

The principal aim of this paper is to establish the correct
theoretical framework to discuss the differences and similarities
between a condensate and a laser, so we adopt a deliberately
simplified model that cannot accommodate all the details of important
physical systems. Nevertheless, in closing we attempt estimates of the
relevant parameters for some relevant physical systems.

To obtained the phase diagram we have considered the following
Hamiltonian: $H = H_0+H_{SB}+H_{B}$, where $H_0 = \omega_c
\psi^\dagger\psi + \sum_{j=1}^N \epsilon_j
(b^\dagger_jb_j-a^\dagger_ja_j) +
\frac{g_j}{\sqrt{N}}(b^\dagger_ja_j\psi + \psi^\dagger
a^\dagger_jb_j)$.  $H_0$ describes an ensemble of N two-level
oscillators with an energy $\epsilon_j$ dipole coupled to one cavity
mode.  $b$ and $a$ are fermionic annihilation operators for an
electron in an upper and lower states respectively and $\psi$ is a
photon bosonic annihilation operator. The sum here is over the
possible sites where an exciton can be present (different molecules or
localised states in the disorder potential). $H_{SB}$ contains all
interactions with an environment:
\begin{multline}
H_{SB}=\sum_{k} g^{\kappa}_{k}(\psi^\dagger d_{k}  +
d_{k}^\dagger \psi)+
\sum_{j,k}[g^{\gamma_{\uparrow}}_{j,k}
(b^\dagger_ja_j c_{j,k}^{\alpha\dagger}+c_{j,k}^{\alpha}
a^\dagger_jb_j)\\+ 
g^{\gamma_{\downarrow}}_{j,k}
(b^\dagger_ja_j c_{j,k}^{\beta}+ 
c_{j,k}^{\beta \dagger}
a^\dagger_jb_j) 
+g^{\gamma_1}_{j,k}(b^\dagger_jb_j
+a^\dagger_ja_j)(c_{j,k}^{\zeta\dagger} +c_{j,k}^\zeta) \\
+g^{\gamma_2}_{j,k}(b^\dagger_jb_j
-a^\dagger_ja_j)(c_{j,k}^{\theta\dagger} +c_{j,k}^\theta)]
\label{eq:HSB}
\end{multline}
and the $H_{B}$ is a Hamiltonian for the baths.  The first term in
(\ref{eq:HSB}) gives rise to the decay of the photon field from the
cavity.  The second term describes an incoherent pumping of two-level
oscillators. The third term contains all the processes which destroy
the electronic excitations such as the decay to photon modes different
to the cavity mode.  These processes, apart from a dephasing effect,
will cause a flow of energy through the system.  However, in the
steady state the total number of excitations, $n_{ex}=
\psi^\dagger\psi + \frac{1}{2}
\sum_{j}(b^\dagger_jb_j-a^\dagger_ja_j)$, which is the sum of photons
and excited two-level oscillators, is conserved.

Finally, the fourth and the fifth terms describe all the dephasing
processes which do not change the total number of excitations in the
cavity, for example collisions and interactions with phonons and
impurities. They can be divided into a part, which acts in the same
way (fifth term) and with an opposite sign (fourth term) on the upper
and lower levels.  These four terms contain all the essential
groups of decoherence processes in the media.

Processes, described by the second, the third and the fourth terms in
(\ref{eq:HSB}), are analogous to the pair-breaking, magnetic
impurities in superconductors and correspond to potentials which vary
rapidly in space (on the lengthscale of excitons) or in time. On the
contrary the fifth term contains interactions which vary very slowly
in space in comparison to the size of excitons and in time and are
analogous to normal, non-magnetic, impurities. They do not have any
pair-breaking effects and lead only to an inhomogeneous broadening of
energies.

For very big pumping and photon decay rates the system would be out of
equilibrium. However, if the system is pumped slowly, with a rate
slower than the thermalisation rate equilibrium assumption can be
justified. In this work we present only the influence of the
decoherence on the system leaving the implications of a
non-equilibrium system for the future work.  This does not put any
restrictions on the strength of the decoherence nor the density of
excitations. The pair-breaking processes described by the fourth term
in (\ref{eq:HSB}) give exactly the same decoherence effects as the
pumping and damping of two-level oscillators preserving equilibrium
for any strength of the interaction.  The ratio between the pumping
and all the damping processes which can be arbitrary big even for
small absolute values of both will be expressed in terms of the
excitation density $\rho_{ex}$.

Usually the baths are averaged out before the exciton - photon
interaction is studied which leads to the well-known quantum
Maxwell-Bloch (Langevin) equations for the photon field and
polarisation. When there is a coherent polarisation this procedure is
not valid ( Langevin equations can be used to study the laser regime
but not the polariton or intermediate regime) \cite{us,me}.  Instead,
we must use the self-consistent Green's function techniques similar to
the Abrikosov and Gor'kov theory of gapless superconductivity
\cite{abrikosov-gorkov}.  Decoherence processes are treated by
perturbation theory about a BCS - like Green function for the
interacting exciton-photon system.  The major difference between our
calculations and the Abrikosov and Gor'kov theory
\cite{abrikosov-gorkov} is that we have two order parameters, the
coherent photon field and polarisation connected through the chemical
potential, set by the excitation density.  We consider two-level
systems in contrast to free propagating electrons model used by
Abrikosov and Gor'kov. Within the Markov approximation we can express
the influence of the bath by a single parameter
$\gamma_l=g^{\gamma_l}(0)^2N_l(0)$ where $l=1,2$ correspond to the
pair-breaking and non-pair breaking decoherence and $N_l$ are
densities of states for the respective baths.  The details of the
method are published elsewhere \cite{us,me}.

\begin{figure}
	\begin{center} 
	\leavevmode 
		\epsfxsize=8.8cm
		\epsfbox{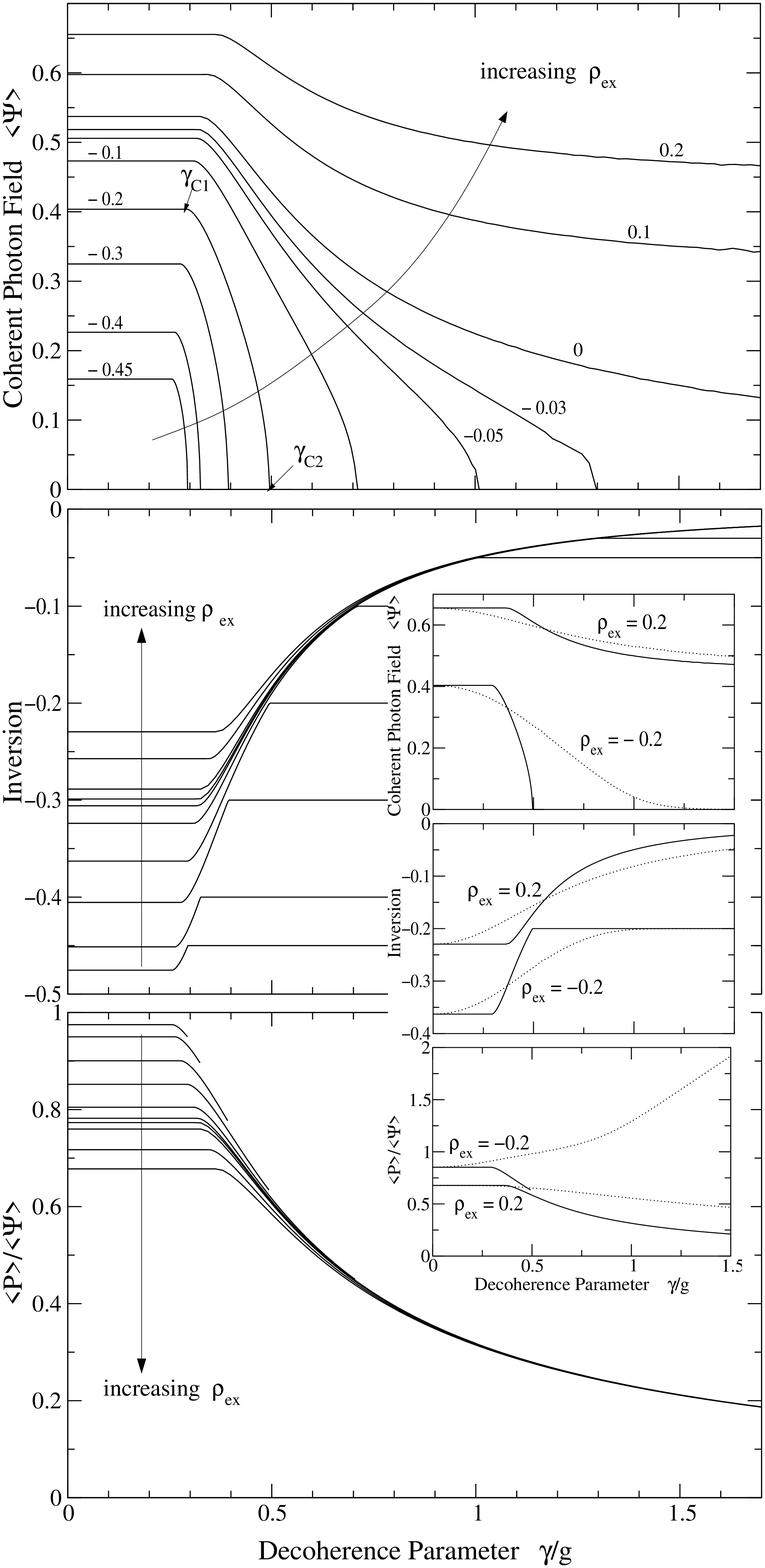}
	\end{center} \caption{Coherent photon field $\langle \psi
	\rangle$ (upper panel), inversion (middle panel) and ratio of
	coherent polarisation and photon field $\langle P
	\rangle/\langle \psi \rangle$ (lower panel) as functions of
	the pair-breaking decoherence strength, $\gamma/g$ for
	different excitation densities, $\rho_{ex}$. The
	coherent fields are rescaled by $\sqrt{N}$ and consequently
	the inversion by $N$. In this terminology the minimum
	$\rho_{ex}=-0.5$ corresponds to no photons and no electronic
	excitations.  Inset: Comparison between the
	influence of a pair-breaking (solid line) and a
	non-pair-breaking (dotted line) decoherence on $\langle \psi
	\rangle$ (upper panel), inversion (middle panel) and $\langle
	P \rangle/\langle \psi \rangle$ (lower panel) for two values
	of $\rho_{ex}$.  }
\label{fig:phot}
\end{figure}

We first determine the ground state properties of the system with
uniform energies, $\epsilon_j=\epsilon$, for different densities in the
presence of the pair-breaking decoherence.  Figure
\ref{fig:phot} shows the ground state behaviour of the coherent fields
and inversion.  For small values of $\gamma/g$, up to some critical
value $\gamma_{C1}$, $\langle \psi\rangle$ is practically unchanged
while for $\gamma/g>\gamma_{C1}$ is damped quite rapidly with the
increasing dephasing (see Fig. \ref{fig:phot} - upper panel). This
critical value of the decoherence strength, $\gamma_{C1}$ is
proportional to $\rho_{ex}$, suggesting that for higher excitation
densities the system is more resistant to the dephasing. At low
excitation densities, where $\rho_{ex}<0$, there is a second critical
value of the decoherence strength, $\gamma_{C2}$, where both coherent
fields are sharply damped to zero.

The ratio of coherent polarisation to coherent field $\langle P
\rangle/\langle \psi \rangle$ (Fig. \ref{fig:phot} - lower panel) for
an isolated system, where $\gamma=0$, depends on the excitation
density.  The condensate becomes more photon like as $\rho_{ex}$ is
increased due to the phase space filling effect. For finite $\gamma$,
at a given excitation density, this ratio decreases with increasing
$\gamma$ meaning that $\langle P\rangle$ is more heavily damped than
$\langle \psi \rangle$.

\begin{figure}
	\begin{center}
	\leavevmode
	\epsfxsize=8.8cm
	\epsfbox{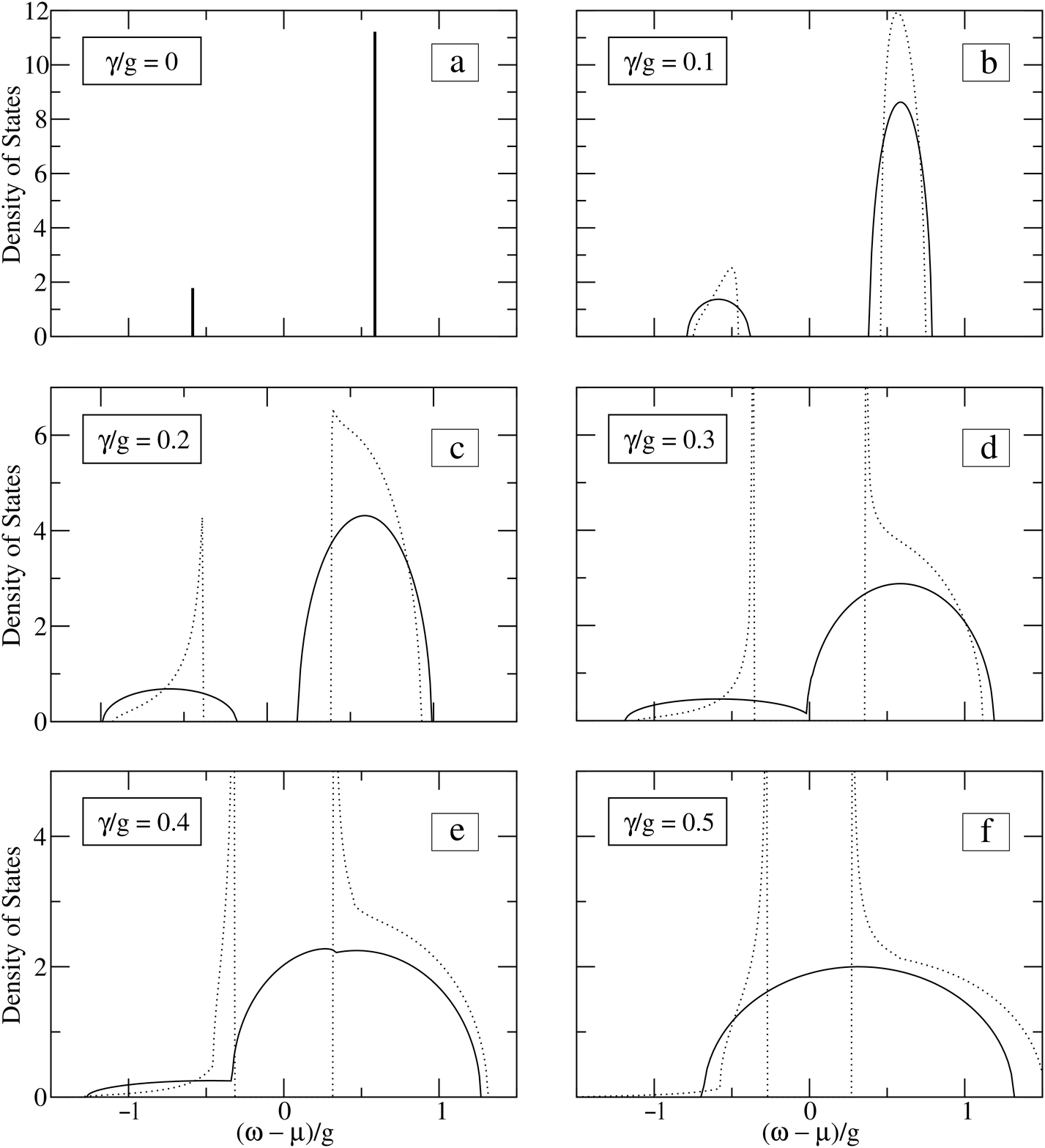}
	\end{center}
	\caption{Density of states for $\rho_{ex}=-0.2$ and different
  	decoherence strengths, $\gamma/g$ for a pair-breaking (solid
  	line) and a non-pair-breaking (dotted line) decoherence
  	processes. f) solid line corresponds to a normal state in which 
  	coherent fields are suppressed.  }
\label{fig:green}
\end{figure}

To understand the behaviour of coherent fields we 
study the excitation spectrum of the system (Fig. \ref{fig:green}).
As $\gamma$ increases the two quasi-particle peaks
(Fig. \ref{fig:green} a) broaden, which causes the decrease in the
magnitude of the energy gap (Fig. \ref{fig:green} b and c).  Finally,
precisely at $\gamma_{C1}$ (shown in Fig.  \ref{fig:phot}), these two
broadened peaks join together and the gap closes (Fig. \ref{fig:green}
d). $\gamma_{C1}$ defines the phase boundary (see Fig \ref{fig:phase})
between a gapped and gapless condensate at low densities and between a
condensate and laser at high densities.

The laser operates in the regime of a very strong decoherence,
comparable with the light-matter interaction itself. In the presence
of such a large decoherence laser action can be observed only for a
sufficiently large excitation density . The coherent polarisation in a
laser system is much more heavily damped than the photon field and a
gap in the density of states is not observed. Thus the laser is a
regime of our system for very large $\rho_{ex}$ and $\gamma$.  It has
to be, however, pointed out that there is no formal distinction
between a laser and a phase-coherent condensate of polaritons. In the
laser, coherence in the media (manifested by the coherent
polarisation) although small, is not completely suppressed so the
laser can be seen as a gapless condensate with a more photon-like
character. One of the possible distinctions between a polariton
condensate and laser could be an existence of an energy gap in the
excitation spectrum.

The dotted lines in onset of Fig. \ref{fig:phot} and in
Fig. \ref{fig:green} show the influence of the non-pair breaking
processes (the fifth term in (\ref{eq:HSB})). We found that these
processes have some quantitative influence on the coherent fields and
the gap in the density of states but do not cause any phase
transitions. Although the two quasiparticle peaks get very broad, the
gap is only slightly affected.  Even for much larger values of
$\gamma$ than presented in Fig. \ref{fig:green} the gap is present
until the coherent fields get completely suppressed. This type of
processes give similar effect as the inhomogeneous broadening of
energy levels.

\begin{figure}
	\begin{center}
	\leavevmode
		\epsfxsize=8.8cm
		\epsfbox{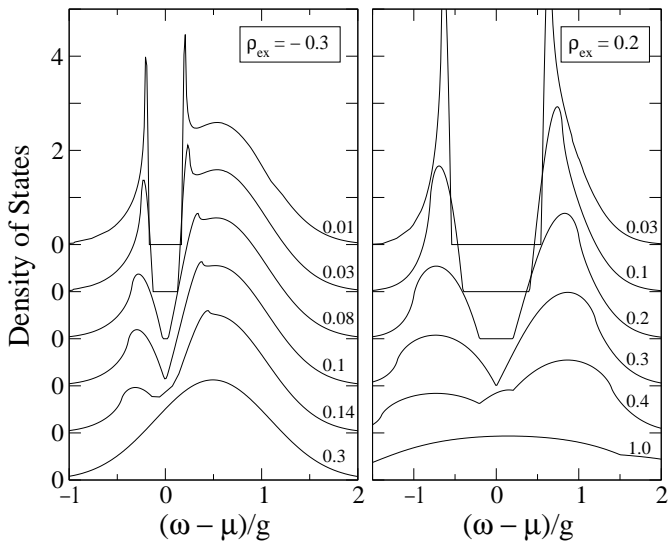}
	\end{center}
	\caption{Density of states for a Gaussian broaden case
	with $\sigma=0.5$ for different values of the pair-breaking
	decoherence at $\rho_{ex}=-0.3$ (left panel) and
	$\rho_{ex}=0.2$ (right panel). }

\label{fig:greenb}
\end{figure}

Finally, we study an influence of the pair-breaking processes on the
system of realistic, inhomogeneously broadened two-level oscillators.
The coherent fields, $\gamma_{C1}$, $\gamma_{C2}$ and the energy gap
are slightly smaller than in the uniform case but all the regimes and
transitions are the same. The broadening of the density of states and
the suppression of the energy gap can be observed as $\gamma$ is
increased (Fig.  \ref{fig:greenb}).

General conclusions about the conditions and systems where polariton
condensation is likely to be observed can be drawn from our theory.
Systems with large dipole coupling $g$, and high densities but low
decoherence are good candidates. Inhomogeneous broadening of
energies have much weaker influence than a homogeneous dephasing.
The large value of $g$ is particularly important as it defines the energy
scale in this problem.  Only the relative strength of the dephasing
and inhomogeneous broadening with respect to $g$ influence
the condensate.  Thus systems with larger $g$ can allow larger values
of decoherence and broadening. The second important factor is the
excitation density. Since the gap is proportional to the amplitude of
the coherent field the condensate is more robust at high
densities. Unfortunately the increase in the density usually implies
an increase in decoherence as the dephasing time ($T_2$) for most of
materials increases linearly with density. An additional increase in
dephasing comes from the increase in the pump intensity required to
achieve higher densities. This could be partially overcome by
improving the quality of the mirrors in microcavity systems.

To be more specific we consider in detail different materials. Among
semiconductors CdTe, ZnSe and GaN are characterised by large values of
$g$ (measured by the polariton splitting in the normal state).  For
CdTe, the $g$ was measured to be 29 meV \cite{g_CdTe} and the
dephasing ($1/T_2$), at the exciton density of 0.05 per Bohr radius, to
be 2.5 meV (0.08 g) \cite{T2_CdTe}. In the absence of inhomogeneous
broadening our theory predicts a gap $\Delta=0.7g=20.3$ meV. In the
case of Gaussian broadening with $\sigma=0.2g=5.8$ meV reported in
some structures the gap is still significant $\Delta=0.08g=2.32$ meV
and it gets suppressed for broadening larger then 8.7 meV.  Other
promising candidates would be excitons in organics, where the
reporting polariton splitting was as large as 80 meV \cite{organic},
or dilute atomic gases in which the dephasing is measured to be around
1000 times smaller then the dipole coupling \cite{atoms} and high
excitations can be achieved without an increase in the density of the
gas.  GaAs --- due to the small value of $g$ --- is not the best
material for the observation of the condensate. Polariton splitting in
GaAs based structures is usually only around 9 meV \cite{g_GaAs} and
the recently reported 20 meV splitting in 36 well structure
\cite{nature} is probably approaching the absolute limit for these
materials.

In this work we have developed a method to include decoherence effects
in the systems with an energy gap in the excitation spectrum, where
the widely used Langevin equations are not valid.  We have studied an
equilibrium, incoherently pumped system. However, in the same way it
would be possible to examine the influence of an environment on
coherently driven condensates.  We have shown that in the presence of
small decoherence the polariton condensate is protected by an energy
gap in the excitation spectrum. The gap is proportional to the
coherent field amplitude and thus the excitation density, so the
condensate is more robust at high densities.  As the decoherence is
increased the gap gets smaller and finally is completely
suppressed. If the polariton condensate is ever to be observed an
increase in density without an increase in decoherence is necessary.
Our method allows us to study different regimes in the cavity and thus
to examine the crossover from a polariton condensate to a laser. The
laser emerges from the polariton condensate at high densities when the
gap in the density of states closes for large decoherence and thus is
analogous to a gapless superconductor. Our work generalises the
existing laser theories to include coherence effects in the
media. When this coherence is included the generation of a coherent
photon field without population inversion becomes possible.

\end{multicols}

\end{document}